\documentclass[aps,pra,twocolumn,tightenlines,amsmath,amssymb,superscriptaddress, 10pt]{revtex4-2}

\usepackage{graphicx}
\usepackage{amsmath}
\usepackage{amssymb}
\usepackage{epstopdf}
\usepackage{color}
\usepackage{lipsum}
\usepackage[colorlinks=true]{hyperref}
\usepackage{ulem}
\usepackage[dvipsnames]{xcolor}
\usepackage{siunitx}
\DeclareGraphicsExtensions{.eps}

\newcommand*{\rom}[1]{\expandafter\@slowromancap\romannumeral #1@}

\begin{document}

\title{Multi-photon emission from a resonantly pumped quantum dot}

\author{F.~Giorgino}
\email{francesco.giorgino@univie.ac.at}
\affiliation{University of Vienna, Faculty of Physics, Vienna Center for Quantum Science and Technology (VCQ), 1090 Vienna, Austria}
\author{P.~Zahálka}
\affiliation{University of Vienna, Faculty of Physics, Vienna Center for Quantum Science and Technology (VCQ), 1090 Vienna, Austria}
\affiliation{Christian Doppler Laboratory for Photonic Quantum Computer, Faculty of Physics, University of Vienna, Vienna, Austria}
\author{L.~Jehle}
\email{lennart.jehle@univie.ac.at}
\affiliation{University of Vienna, Faculty of Physics, Vienna Center for Quantum Science and Technology (VCQ), 1090 Vienna, Austria}
\author{L.~Carosini}
\affiliation{University of Vienna, Faculty of Physics, Vienna Center for Quantum Science and Technology (VCQ), 1090 Vienna, Austria}
\affiliation{Christian Doppler Laboratory for Photonic Quantum Computer, Faculty of Physics, University of Vienna, Vienna, Austria}
\author{L.~M.~Hansen}
\affiliation{University of Vienna, Faculty of Physics, Vienna Center for Quantum Science and Technology (VCQ), 1090 Vienna, Austria}
\affiliation{Christian Doppler Laboratory for Photonic Quantum Computer, Faculty of Physics, University of Vienna, Vienna, Austria}
\author{J. C. Loredo}
\altaffiliation[Current address: ]{Sparrow Quantum, Blegdamsvej 104A, 2100 Copenhagen, Denmark}
\affiliation{University of Vienna, Faculty of Physics, Vienna Center for Quantum Science and Technology (VCQ), 1090 Vienna, Austria}
\affiliation{Christian Doppler Laboratory for Photonic Quantum Computer, Faculty of Physics, University of Vienna, Vienna, Austria}
\author{P.~Walther}
\affiliation{University of Vienna, Faculty of Physics, Vienna Center for Quantum Science and Technology (VCQ), 1090 Vienna, Austria}
\affiliation{Christian Doppler Laboratory for Photonic Quantum Computer, Faculty of Physics, University of Vienna, Vienna, Austria}
\affiliation{ University of Vienna, Research Network for Quantum Aspects of Space Time (TURIS), 1090 Vienna, Austria}
\affiliation{Institute for Quantum Optics and Quantum Information (IQOQI) Vienna, Austrian Academy of Sciences, Vienna, Austria}

\begin{abstract}
Resonance fluorescence of natural or artificial atoms constitutes a prime method for generating non-classical light.
While most efforts have focused on producing single-photons, multi-photon emission is unavoidably present in the resonant driving of an atom.
Here, we study the extent to which these processes occur: we quantify the multi-photon emission statistics in a resonantly-driven two-level artificial atom---a semiconductor quantum dot in a micropillar cavity---when pumped by short optical pulses.
By measuring auto-correlation functions up to the fourth order, we observe up to four photons emitted after a single pumping pulse, and investigate the emission dynamics with finely resolved temporal measurements.
Furthermore, we propose a method based on acquisition time gating to enhance the purity of a single-photon source while maintaining high efficiencies.
Our results deepen the understanding of the photon emission processes in coherently driven atomic systems, and suggest a simple but effective technique to reduce the multi-photon components of a single-photon source.
\end{abstract}

\maketitle

\section{\label{sec:intro} Introduction}

The driving of atomic transitions is widely employed in the creation of non-classical states of light, demonstrated across various platforms including atoms~\cite{Kuhn2002, Beugnon2006}, molecules~\cite{Toninelli2021}, and solid-state systems~\cite{Aharonovich2016}.
Out of a plethora of atomic systems, semiconductor quantum dots stand out due to their exceptional optical properties~\cite{Tomm2021, Liu2019}, combined with the scalability offered by solid-state technology, facilitating seamless integration into photonic quantum hardware~\cite{Wang2023}. Among several excitation methods, resonance fluorescence is known for generating quantum light with high degrees of source efficiency, indistinguishability, and coherence~\cite{ding2023, Tomm2021, Loredo2019}.
This makes the intricate interaction between light and matter in quantum dots akin to atomic experiments.

In an ideal two-level system (TLS), oscillations between the ground state and the excited state occur as a function of the absorbed resonant pulse area $\Theta$.
However, in realistic TLSs, spontaneous emission plays a significant role for the atom-photon dynamics.
During spontaneous emission, the excited state relaxes back to the ground state, releasing energy in the form of light.
This natural process has been extensively utilized to create high-quality triggered non-classical light sources. 
Here, laser pulses much shorter than the radiative lifetime of the TLS excite the atomic system while reducing spontaneous emission during the system-pulse interaction.
Nevertheless, driving the TLS by pump pulses with areas of $\Theta{=}2n\pi$, where $n$ is an integer, results in a significant probability for a two-photon emission~\cite{Fischer2017}. Whenever a spontaneous decay occurs during the pump-system interaction, there is a chance to re-excite the TLS causing the emission of a second photon.
Limited to two-photon events, this process has been described before \cite{Fischer2017, Fischer2017_2} but the generation of higher multi-photon events during the resonant driving of realistic atoms has not yet been fully studied.
Nonetheless, a comprehensive understanding of the underlying dynamics is necessary to generate more complex quantum states using atomic systems.

\begin{figure}[b!]
    \begin{center}
    \includegraphics[width=0.5\textwidth]{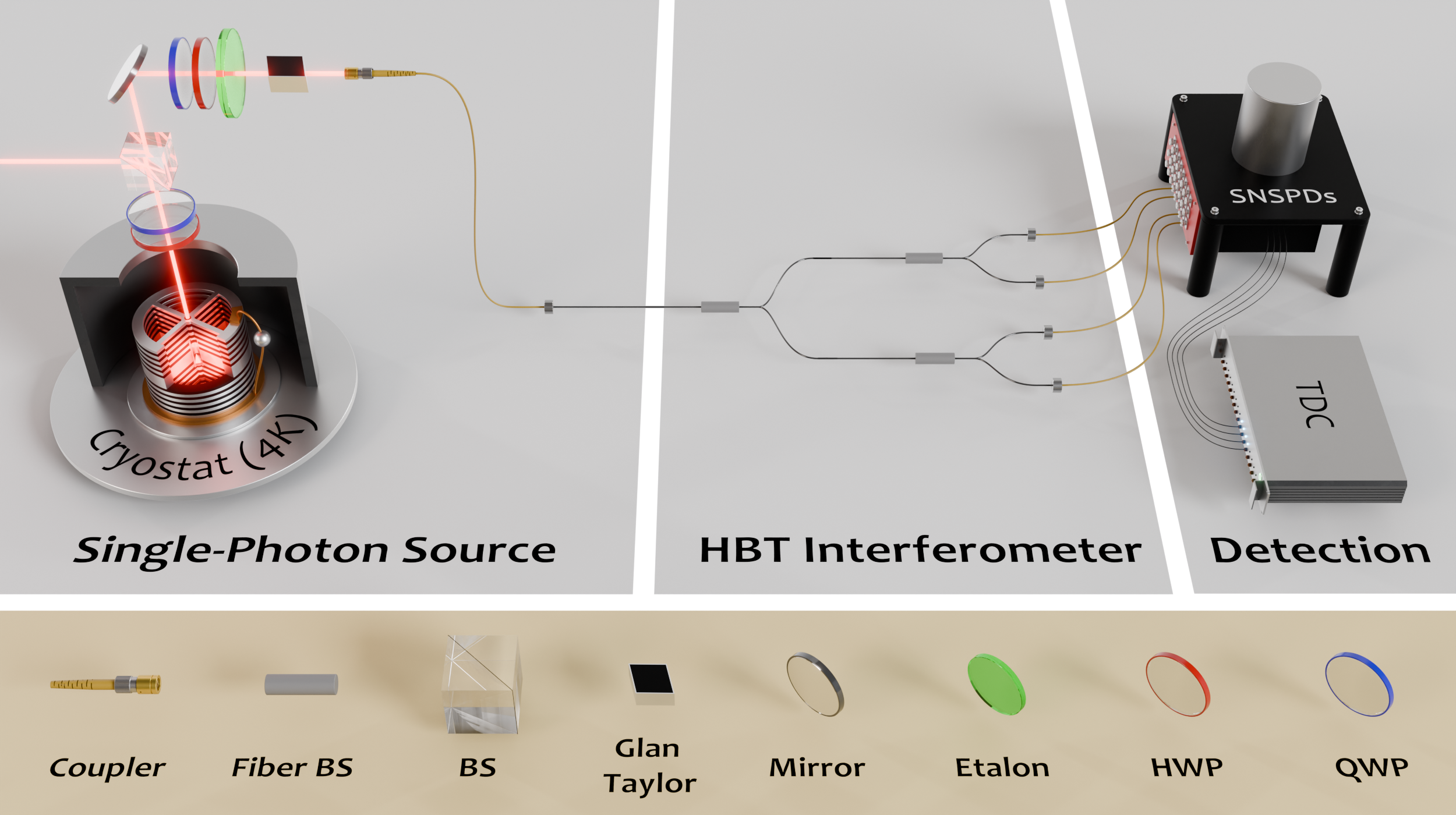}\vspace{0mm}	
    \caption{{\bf Generalised Hanbury-Brown Twiss experiment.} The photons from the source are probabilistically split in four paths and are detected with Superconducting Nanowires Single Photon Detectors (SNSPDs). The clicks are recorded and correlated with a tagging logic, simultaneously building cross-correlation histograms.}
    \label{fig:concept}
    \end{center}
\end{figure}

\begin{figure*}[t!]
    \centering
    \includegraphics[width=\textwidth]{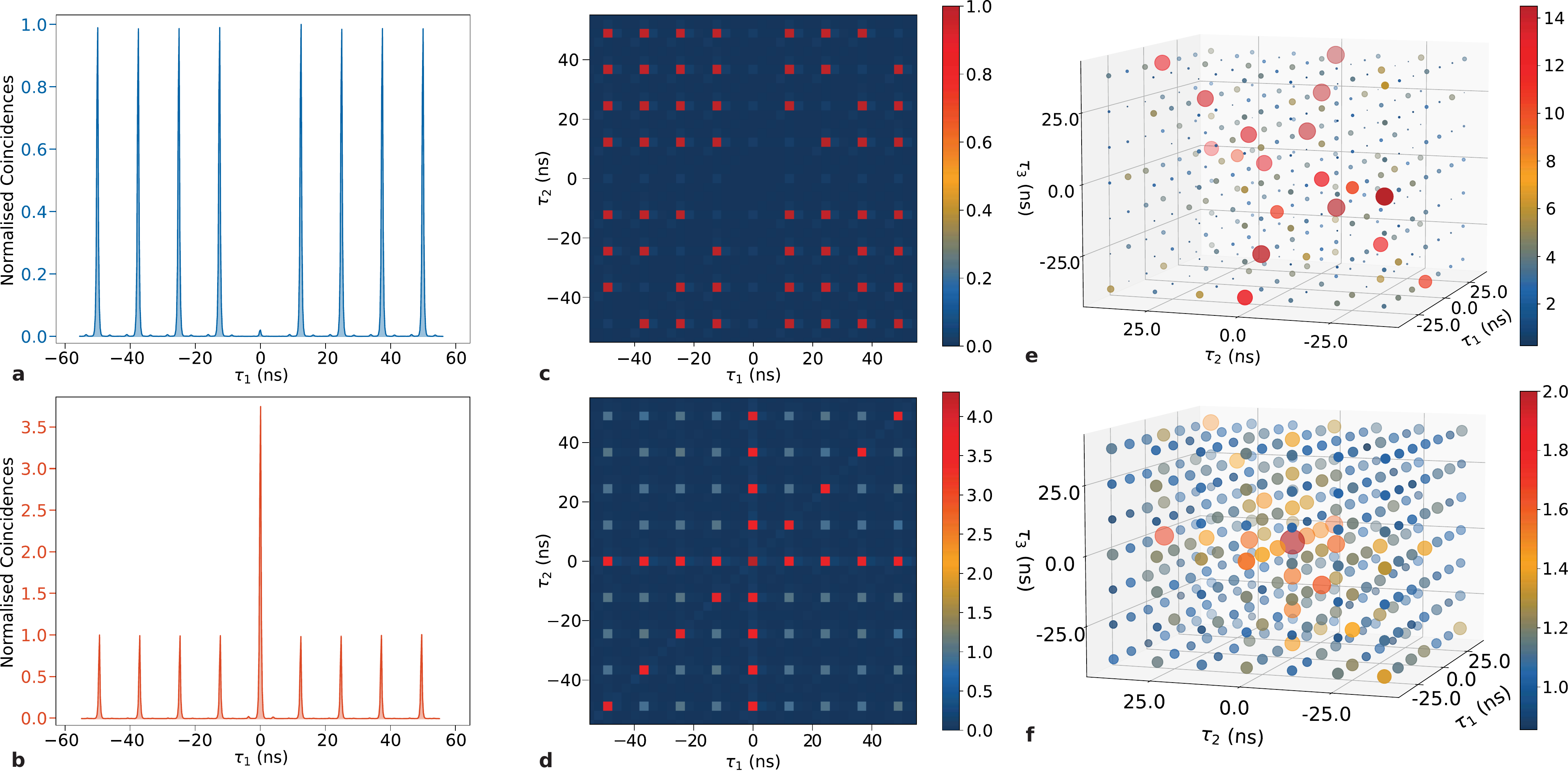}\vspace{0mm}
    \caption{{\bf Auto-correlation histograms.} {\bf a - b} Second-order auto-correlation function $g^{(2)}(\tau_1)$  under a $\pi-$ and $2\pi-$ pulse, showing anti-bunching  (\textbf{a}) $g^{(2)}(0) = 0.031 \pm 0.001$ and bunching (\textbf{b}) $g^{(2)}(0) = 4.08 \pm 0.01$. {\bf c - d} Analogous correlation maps $g^{(3)}(\tau_1, \tau_2)$  yielding $g^{(3)}(0, 0) = (5.08^{+0.9}_{-1})\cdot 10^{-4} $ at $\pi$ pulse (\textbf{c})  and $g^{(3)}(0, 0) = 4.31 \pm 0.08$ at $2\pi$ (\textbf{d}). The map has been built with a bin width of $\SI{3.125}{ns}$. {\bf e - f}  Normalized fourth-order auto-correlation function $g^{(4)}(\tau_1, \tau_2, \tau_3)$ for $\Theta = 2\pi$ and $6\pi$. The histograms are built binning the raw data with a window of $\SI{12.5}{ns}$. To guide the eye, the value of each peak is represented both by color and size of the plotted data point.}
    \label{fig:auto-correlations}
\end{figure*}

Here, we investigate the characteristics of multi-photon states emitted from a TLS when driven by resonant optical pulses---a trion transition in a quantum dot (QD) embedded in a micropillar cavity.
In particular, we study features of multi-photon emissions for pump areas beyond $6{\pi}$-pulses, and measure auto-correlation functions up to the fourth order; consequently, quantifying the photon-number emission probabilities up to four photons.
Moreover, we explore the time-dependent behavior of these multi-photon statistics, and demonstrate a time-gated filtering technique that allows obtaining higher single-photon purities---without strong compromise on the source efficiency---as well as higher photon indistinguishabilities.

\begin{figure*}[t!]
    \centering
    \includegraphics[width=\textwidth]{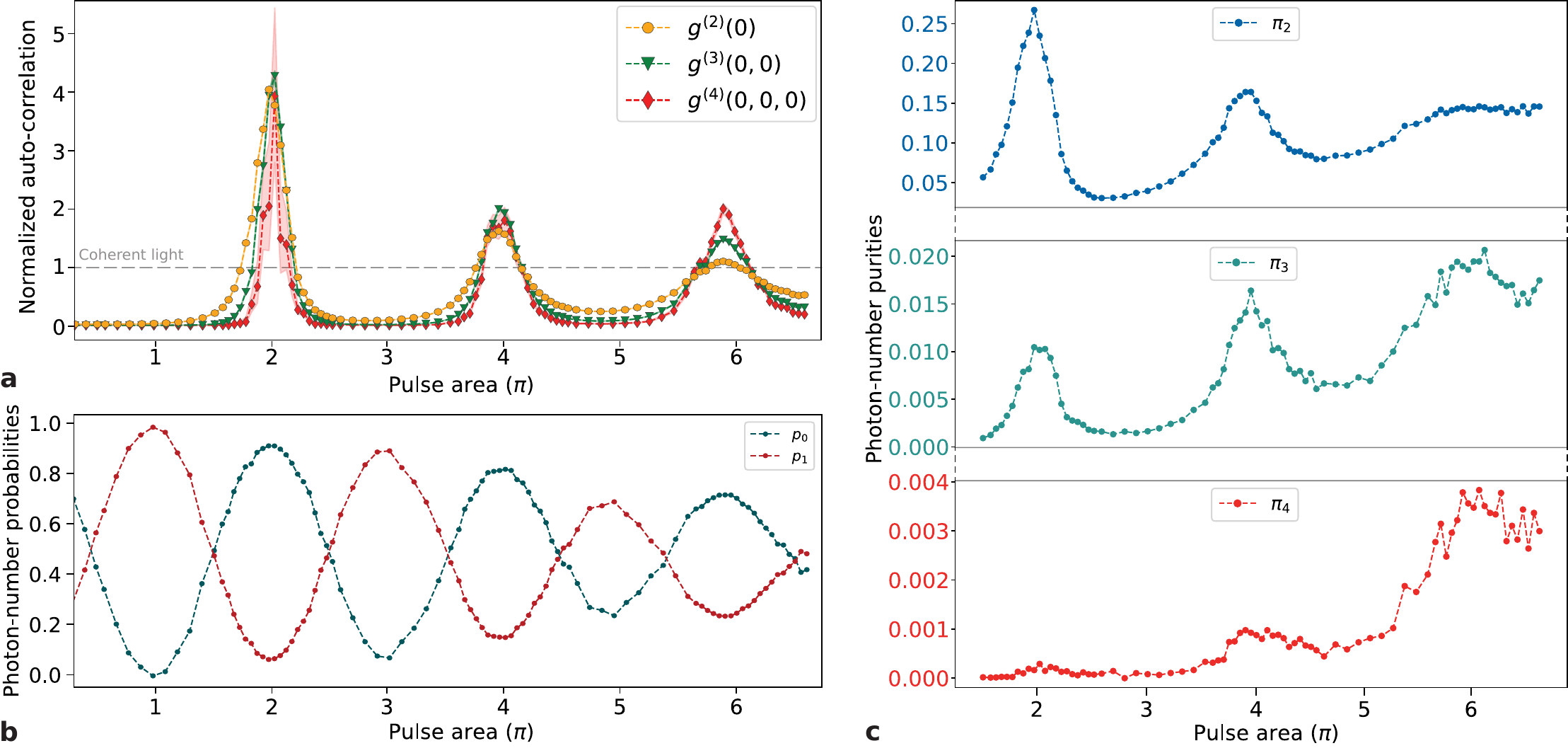}\vspace{0mm}	
    \caption{ {\bf Photon-number statistics }{\bf a}
    Experimental auto-correlation measurements versus pulse area showing oscillations between sub-Poissonian statistics (at odd $\pi-$pulses) and super-Poissonian statistics (at even $\pi-$pulses). The shaded areas show $1\sigma$ confidence intervals of the measurement, calculated assuming Poissonian counting statistics. {\bf b} Probability of emitting zero or one photon versus excitation pulse area, showing clear Rabi rotations up to $6\pi$. The oscillations are damped by the phonon-induced dephasing mechanism that increases with the magnitude of the pumping pulse. {\bf c} Two-, three- and four-photon emission probabilities for different pulse areas, renormalized by excluding the vacuum probability. The lines exhibit similar trends with significant differences in magnitude, for instance at $\Theta = 2\pi$: $\pi_2 \sim 27\%$, $\pi_3 \sim 1\%$, and $\pi_4 \sim 0.05\%$.}
    \label{fig:pn-probabilities}
 \end{figure*}

\section{Results}
\subsection{\label{sec:auto-correlation} Auto-correlation functions}

The second-order auto-correlation function at zero time delay, denoted as $g^{(2)}(0)$, serves as a common metric to assess the purity of a single-photon source \cite{Glauber1963}.
A value approaching 0 indicates a negligible presence of multi-photon states.
In experiments, $g^{(2)}(0)$ is determined using a Hanbury-Brown and Twiss (HBT) setup, measuring coincidence counts between two single-photon detectors at the output of a balanced beam splitter \cite{Brown1956}.
Our generalized HBT experiment, sketched in Fig. \ref{fig:concept}, involves recording up to four-photon coincidences and estimating $g^{(m)}(\mathbf{\tau})$ with $m={2,3,4}$ while varying the pumping power.
We denote $\mathbf{\tau} = (\tau_1, ..., \tau_{m-1})$ as the vector of time delays between different channels of the HBT setup. 

Under $\pi$-pulse excitation, the auto-correlation histograms, see Figs. \ref{fig:auto-correlations}a,c, exhibit strong anti-bunching, $g^{(2)} (0) < 1$, and show $g^{(3)} (0, 0) < g^{(2)} (0)$ as expected from a single-photon source where three-photon coincidences are significantly fewer than the two-photon events.
However, when probed with $2\pi$-pulses, the system shows bunching statistics (see Figs. \ref{fig:auto-correlations}b,d) witnessing the presence of multi-photon bundles \cite{Muoz2014, Fischer2017}.

Analyzing the higher-order correlation functions not just at zero time delay but presenting them as function of $\tau$ unveils intricate details. 
For instance, the peaks along the lines of $\tau_1 = 0$, $\tau_2 = 0$, or $\tau_1 = \tau_2$ in Fig. \ref{fig:auto-correlations}c,d correspond to two detectors clicking simultaneously, that is, those points estimate the $g^{(2)}(0)$.
Furthermore, the four-photon coincidence histogram displays the magnitude of the different multi-photon correlations.
At $\Theta = 2\pi$ (see Fig. \ref{fig:auto-correlations}e), the histogram shows prominent peaks along three lines, that is when the four detectors click pairwise simultaneously, which estimates the $[g^{(2)} (0)]^2$. At $\Theta\,=\,6\pi$ (see  Fig. \ref{fig:auto-correlations}f), a distinct pattern emerges, with the central peak---the $g^{(4)}(0, 0, 0)$---dominating over two or three detectors firing jointly, indicating $g^{(4)}(0, 0, 0) > g^{(3)}(0, 0) > g^{(2)}(0)$.  

Evaluating $g^{(m)}(0)$ as a function of pulse area, shown Fig. \ref{fig:pn-probabilities}a, clearly demonstrates oscillations between sub-Poissonian statistics around odd-$\pi$ areas and super-Poissonian around even-$\pi$ areas.
Whenever the pumping power is close to drive an even number of Rabi cycles, i.e. $\Theta = 2n\pi$, the emitted field exhibits a strong multi-photon component resulting in the observed bunching behavior.
The emission of multi-photon states is attributed to the re-excitation of the QD during the excitation pulse duration, as discussed in \cite{Fischer2017}.
For resonance fluorescence, this process has been shown to scale linearly with pulse duration \cite{Hanschke.2018} and is predicted even for arbitrarily short pumping pulses.

Although auto-correlations provide insights into the non-classicality, a thorough---and quantitative---understanding of the photon emission statistics produced by the TLS must be derived from the photon-number probabilities.
Here, we refer to the photon-number probability $p_n$ as the probability of emitting $n$ photons as consequence of a single excitation pulse, irrespective of their individual temporal ordering..
In Sec. \ref{sec:emission_dinamics}, we deepen the analysis by resolving the temporal evolution of the emission statistics. 
Based on our findings, we argue that, even for a high $p_2$, QDs do not produce two-photon Fock states but emit the two photons in different temporal modes.

\begin{figure*}[t!]
    \centering
    \includegraphics[width=\textwidth]{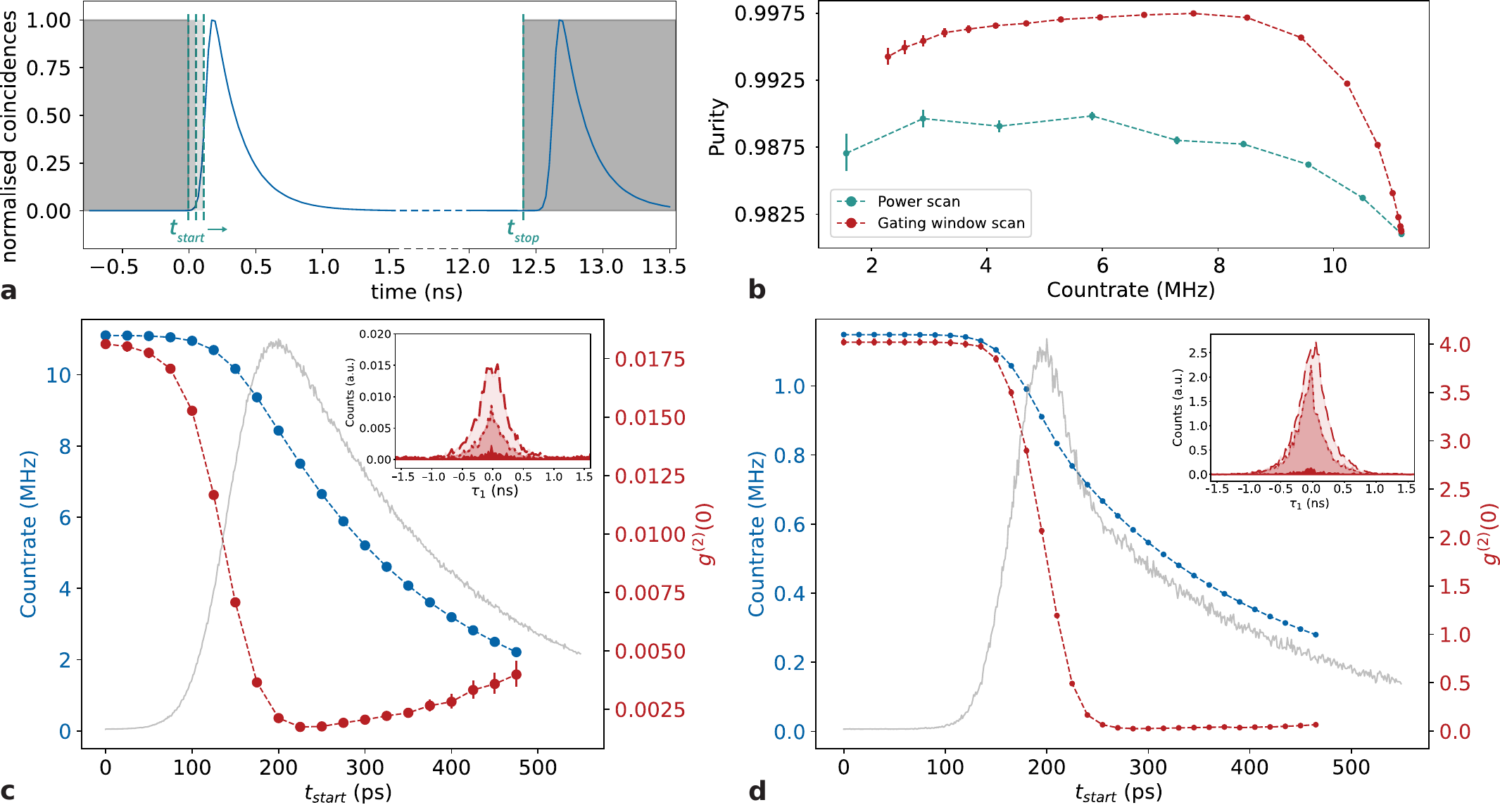}\vspace{0mm}	
    \caption{ {\bf Study of the emission dynamics} {\bf a} Scheme of the gated measurement. The signals have been appropriately delayed so that the lifetime peak is always $\SI{200}{ps}$ shifted from the reference clock. The beginning of the gating window $t_{start}$ is determined by the clock and it is moved forward, in order to cut an increasing portion of the incoming signal. The end of the gating window $t_{stop}$ is fixed and corresponds to the clock signal delayed by $\SI{12.4}{ns}$. Only the events falling in this window are accounted for in the $g^{(2)}(0)$ evaluation. {\bf b} Behavior of single-photon purity $P= 1 - g^{(2)}(0)$ as a function of the measured count rate, for different excitation intensities -- pulse area ranging from $0.1\pi$ to $\pi$ -- and for different gating windows at $\pi$-pulse excitation. It shows that gating the acquisition yields higher purity as compared to decreasing the excitation power. {\bf c - d} Count rate and $g^{(2)}(0)$ as a function of $t_{start}$, for $\pi-$pulse on the left plot and $2\pi-$pulse on the right one. The auto-correlation decreases substantially when the rising edge of the photons lifetime, displayed in Grey in the background, is excluded by acquisition gating. 
    The insets show the progressive reduction of the normalized peak at zero time delay when $t_{start} = 0 / 150 / \SI{225}{ps}$ ($0 / 200 / \SI{255}{ps}$) for a $\pi (2\pi )$ -pulse area.
    Note that in ({\bf c}) $g^{(2)}(0)$, after reaching a minimum, increases again as a result of the decreasing signal-to-noise ratio, as no background subtraction is performed on the data. This behavior is not visible in ({\bf d}) because of the different y-axis scaling.}
    \label{fig:emission_dinamics}
\end{figure*}

\subsection{\label{sec:pn-probabilities} Photon-number probabilities}
While we are interested in the photon-number statistics emitted from the source, denoted as ${p_n}$, experimentally, only the photon-number distribution measured at the detectors, ${p_n'}$, are accessible. The coherence functions $g^{(m)}(0)$ depend on this probability distribution, as described by
\begin{equation}
    g^{(m)} (0) = \frac{\sum_n n(n-1) ... (n-m+1)) \ p'_n }{\left[\sum_n n \ p'_n\right]^n}  \ .
\end{equation}
In our analysis, we neglect contributions beyond the fourth order, setting $p_{n>4} = 0$.
Taking also the normalization into account, $\sum_{n=0}^4 p_n' = 1$, we can thus compute $p_n'$ up to $n=4$ using the measurements of $g^{(m)}(0)$ for $m = {2,3,4}$ and the source efficiency measured at the detector, $B' = \sum_{n=1}^4 p_n'$.
The later is obtained by dividing the measured count rate by the repetition rate of the excitation laser.
Since any loss distorts the distribution (higher photon-number probabilities are suppressed), we compute ${p_n}$ from the measured statistics ${p_n'}$ by correcting for $\eta_\text{t}$, a parameter that encapsulates all the linear losses experienced by the wave-packet after the emission from the atom.
Assuming that a TLS driven by a $\pi$-pulse emits at least one photon we estimate this transmission probability to be $\eta_{\text{t}}\approx\SI{25}{\percent}$, and retrieve ${p_n}$ (see Supplementary Material).

As expected, Fig. \ref{fig:pn-probabilities}b presents the intensity-damped Rabi rotations characteristic for a TLS.
Comparing to Fig. \ref{fig:pn-probabilities}a, we note that whenever bunching statistics are observed---at $\Theta = 2n\pi$---also the vacuum contribution $p_0$ is maximized. To analyze the features of the higher photon-number states more clearly, we remove the effect of $p_0$ by using the photon-number purities
\begin{equation}
    \pi_n = \frac{p_n}{\sum_{n>0} p_n} \ 
\end{equation}
as a renormalization of the populations. 
As shown in Fig.~\ref{fig:pn-probabilities}c, the probabilities of producing multi-photon states follow the same trend as the bunching behavior described by the auto-correlation functions.
Remarkably, however, a bunching $g^{(2)}$ does not imply that the multi-photon component is larger than the single-photon probability.
In fact, at $\Theta = 2\pi$, despite the strong bunching shown in Fig. \ref{fig:auto-correlations}b, we observe only $\pi_2 \approx 27\%$.
These results are explained by the impact of the vacuum component on the auto-correlation measurements. Specifically, the condition $g^{(2)}(0)>1$ can only be met if the vacuum probability is sufficiently high, regardless of the values of $p_{n>0}$. See Supplementary Material for a more detailed analysis.

\begin{figure}[t]
    \begin{center}
    \includegraphics[width=0.5\textwidth]{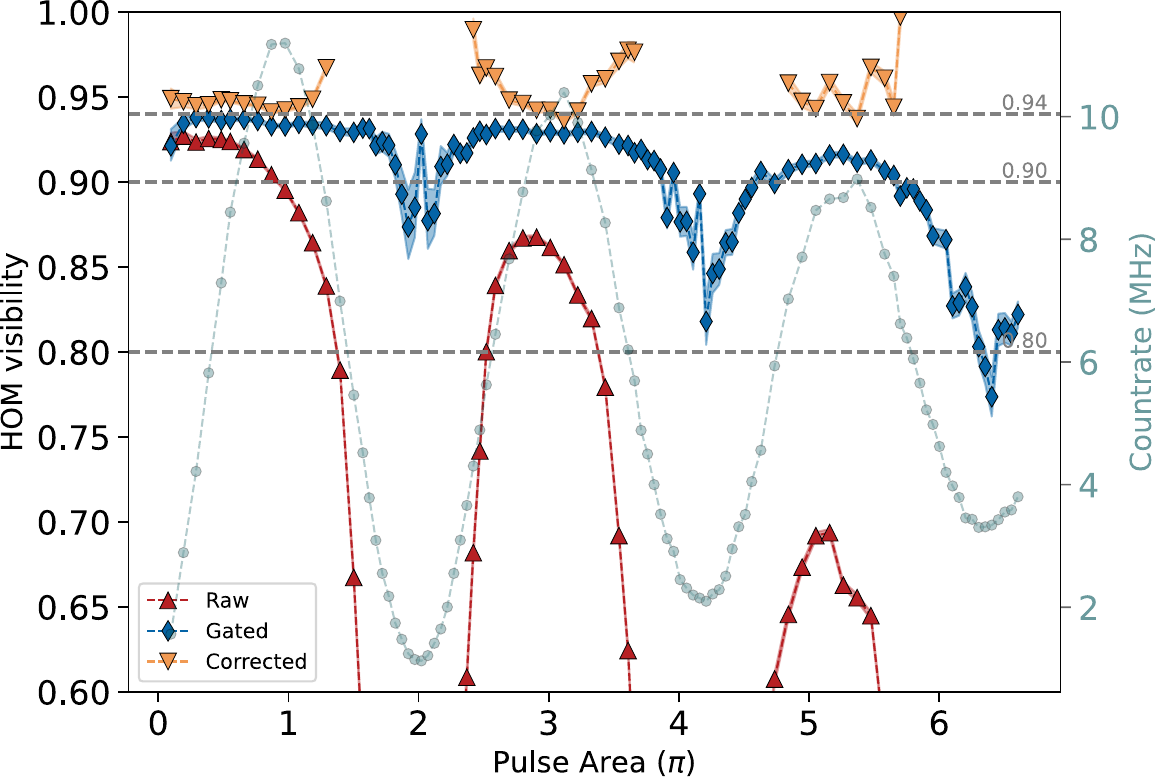}\vspace{0mm}	
    \caption{{\bf Hong-Ou-Mandel interference.} Correlating the output of an unbalanced MZI extracts the raw visibility $V_{\mathrm{HOM}}$ shown by the red triangles. Using the formula in \cite{Ollivier2021}, we extract the mean wave packet overlap $M$ correcting the visibility for the multi-photon component which has been measured during the same acquisition along with the count rate (displayed in the background). 
    Only the data points close to odd multiples of $\pi$ are displayed, since the standard HOM setup only assesses the indistinguishability of single photons. 
    We additionally performed a gated analysis where portion of the incoming signal is neglected, maximizing the raw visibility of the remaining wave-packet $V_{\mathrm{HOM}}^{\mathrm{g}}$.
    }
    \label{fig:hom}
    \end{center}
\end{figure}

\subsection{\label{sec:emission_dinamics} Emission Dynamics and indistinguishability}

The full understanding of the multi-photon emission process requires a study of the time dynamics of the system.
As sketched in Fig. \ref{fig:emission_dinamics}a, we evaluated the second-order auto-correlation function $g^{(2)}(0)$ on a selected portion of the incoming signal.
Fig. \ref{fig:emission_dinamics}c-d shows that, disregarding the portion of photons arriving early, that is when the laser is still interacting with the system, results in a decreasing $g^{(2)}(0)$.
This measurement confirms that multi-photon events are caused by multiple emissions happening in subsequent time bins and shows that acquisition gating reduces the multi-photon components over-proportionally compared to single-photon events. 

Neglecting the initial portion of emitted signals is hence useful to improve the purity of single-photon sources based on two-level systems,  while maintaining high brightness.
For instance, see Fig. \ref{fig:emission_dinamics}c, with $t_{start} = \SI{150}{ps}$ the purity increases by a factor of $\sim60\%$ while the count rate is only reduced by a factor of $\sim8.5\%$.  
We note that, in the case of resonant excitation, this method is favorable compared to power tuning as shown in Fig. \ref{fig:emission_dinamics}b, where, for equal rates, the gating provides higher levels of purity. This approach is effective due to the temporal structure of the multi-photon wave -packets where the first $n-1$ photons — emitted while the laser is still present — is temporally shorter, while the last follows the characteristic decay of an isolated two-level system. This is particularly evident in Fig. \ref{fig:emission_dinamics}d, where the significant portion of two-photon states modifies the decay rate.

We further conduct a Hong-Ou Mandel (HOM) interference experiment \cite{Hong1987}, assessing the source indistinguishability under different pumping powers.
Clicks at the output of an unbalanced Mach-Zehnder Interferometer (MZI) are recorded for each pulse area, from which the raw visibility $V_{\mathrm{HOM}}$ is computed.
Employing the methodology outlined in \cite{Ollivier2021}, we derive the mean wave packet overlap $M$, adjusting the visibility $V_{\mathrm{HOM}}$ to account for multi-photon contributions, quantified by the second-order intensity auto-correlation at zero time delay $g^{(2)}(0)$.

However, note that a HOM experiment aims at assessing the indistinguishability of a single photon source.
Hence, the raw and corrected visibility data points depicted in Fig. \ref{fig:hom} offer meaningful insights in regions where the pumping pulse area is an odd multiple of $\pi$.
Notably, despite increasing multi-photon contributions compromising the HOM visibility as $\Theta$ grows, the wave-packet remains highly indistinguishable, with $M \sim 94\%$. 

We then evaluate the HOM visibility again, but using the gated acquisition, similar to Fig. \ref{fig:emission_dinamics}a, where a portion of the incoming signal is disregarded.
Gating removes the multi-photon component stemming from re-excitation.
Consequently, for $\Theta = (2n+1)\pi$ the gated visibility $V_{\mathrm{HOM}}^{\mathrm{g}}$ closely approximates the wave-packet overlap $M$.
However, the discrepancy between these measures widens with increasing pulse area as other sources of multi-photon emission, such as increasing phonon sidebands, gain prominence.  

For $\Theta = 2n\pi$, neglecting photons emitted during the system-pump interaction, allows for a rough estimation of source indistinguishability.
However, a significant challenge arises from the predominant vacuum state within the output field, in fact, filtering out a large portion of the signal leads to a significant background contribution, which in this work is not subtracted.



\section{\label{sec:discussion} Discussion}

While conceptually similar investigations have been reported for  laser, thermal or QD sources \cite{Dynes2018, Amann2009, Rundquist2014, Stevens2014, Fischer2017}, each focused only on a particular aspect of the number statistics, including photon bunching \cite{Fischer2017} or $g^3(0)$ measurements \cite{Stevens2014}.

In contrast, we presented comprehensive experimental data from a resonantly driven QD single-photon source, showcasing both its emission statistics and dynamics in great detail.
The high efficiency and controllability of our system enabled measurements of the auto-correlation functions up to fourth order and record the evolution for pulses areas beyond $6\pi$.
The high-order histograms reveal a complex behavior, even for a system as simple as a TLS.
Inferring the photon-number purities from the histograms, we showed that all multi-photon contributions, $\pi_2, \pi_3, \pi_4$, are maximized close to even pulse areas, and the importance of the vacuum component $p_0$ to observe bunching statistics was highlighted.

Furthermore, based on finely time-resolved measurements, we explore the QD dynamics causing the multi-photon events and confirm that the photons do not share the same temporal mode but stem from successive spontaneous emissions. We conclude that the emission from a resonantly-driven 2TL even at large pulse areas is strictly time-ordered and does not exhibit higher-order Fock states.
Finally, we demonstrate how insights in temporal shape can be leveraged to improve the single-photon purity by acquisition time gating. 

\section{\label{sec:acknowledgement} Acknowledgement}

This research was funded in whole, or in part, by Horizon 2020 and Horizon Europe research and innovation programme under grant agreement No 101135288 (EPIQUE), the Marie Skłodowska-Curie grant agreement No 956071 (AppQInfo), and the QuantERA II Programme under Grant Agreement No 101017733 (PhoMemtor). Further funding was received from the Austrian Science Fund (FWF) through Quantum Science Austria (COE1), BeyondC (Grant-DOI 10.55776/F71) and Research Group 5(FG5); and from the Austrian Federal Ministry for Digital and Economic Affairs, the National Foundation for Research, Technology and Development and the Christian Doppler Research Association. For the purpose of open access, the author has applied a CC BY public copyright licence to any Author Accepted Manuscript version arising from this submission.

\bibliography{multiphoton}

\end{document}


\title{Supplementary Material: Multi-photon emission from a resonantly pumped quantum dot}

\author{F. Giorgino}
\email{francesco.giorgino@univie.ac.at}
\affiliation{University of Vienna, Faculty of Physics, Vienna Center for Quantum Science and Technology (VCQ), 1090 Vienna, Austria}
\author{P. Zahálka}
\affiliation{University of Vienna, Faculty of Physics, Vienna Center for Quantum Science and Technology (VCQ), 1090 Vienna, Austria}
\affiliation{Christian Doppler Laboratory for Photonic Quantum Computer, Faculty of Physics, University of Vienna, Vienna, Austria}
\author{L.~Jehle}
\email{lennart.jehle@univie.ac.at}
\affiliation{University of Vienna, Faculty of Physics, Vienna Center for Quantum Science and Technology (VCQ), 1090 Vienna, Austria}
\author{L.~Carosini}
\affiliation{University of Vienna, Faculty of Physics, Vienna Center for Quantum Science and Technology (VCQ), 1090 Vienna, Austria}
\affiliation{Christian Doppler Laboratory for Photonic Quantum Computer, Faculty of Physics, University of Vienna, Vienna, Austria}
\author{L.~M.~Hansen}
\affiliation{University of Vienna, Faculty of Physics, Vienna Center for Quantum Science and Technology (VCQ), 1090 Vienna, Austria}
\affiliation{Christian Doppler Laboratory for Photonic Quantum Computer, Faculty of Physics, University of Vienna, Vienna, Austria}

\author{J. C. Loredo}
\altaffiliation[Current address: ]{Sparrow Quantum, Blegdamsvej 104A, 2100 Copenhagen, Denmark}
\affiliation{University of Vienna, Faculty of Physics, Vienna Center for Quantum Science and Technology (VCQ), 1090 Vienna, Austria}
\affiliation{Christian Doppler Laboratory for Photonic Quantum Computer, Faculty of Physics, University of Vienna, Vienna, Austria}
\author{P.~Walther}
\affiliation{University of Vienna, Faculty of Physics, Vienna Center for Quantum Science and Technology (VCQ), 1090 Vienna, Austria}
\affiliation{Christian Doppler Laboratory for Photonic Quantum Computer, Faculty of Physics, University of Vienna, Vienna, Austria}
\affiliation{ University of Vienna, Research Network for Quantum Aspects of Space Time (TURIS), 1090 Vienna, Austria}
\affiliation{Institute for Quantum Optics and Quantum Information (IQOQI) Vienna, Austrian Academy of Sciences, Vienna, Austria}

\maketitle
\onecolumngrid

\section{Photon Source}
We employ a charged exciton of an InGaAs quantum dot (QD) embedded in an electrically-tunable micropillar cavity in a sample cooled to $\sim  \SI{4}{K}$ in a low-vibration closed-cycle cryostat.\newline
The $\SI{80}{MHz}$ Fourier-transform limited, pulsed laser is tuned to match the QD-cavity wavelength of $\SI{922.3}{nm}$.
The pump spectrum is shaped with a 4-f system, setting a bandwidth between $100 - \SI{400}{pm}$, therefore obtaining pulse with a duration between $\sim\SIrange{3}{15}{ps}$. \newline
A confocal microscope setup is used to excite the QD and collect its resonance fluorescence signal.
In particular, we place a $97:3$ (transmission/reflection) beamsplitter (BS) and use the reflection mode for excitation, and the transmission mode for collection.
The laser pump is separated from the QD signal in a standard crossed-polarization scheme.
A set of quarter and a half wave plates, $Q_1$ and $H_1$, initialize the laser polarization in the excitation path to one of the linear cavity modes. In the collection path, another set of wave plates, $Q_2$§ and $H_2$, and a Glan-Taylor polarizer suppress the laser with $>OD7$.
A $\SI{70}{\pm}$ bandwith etalon filter (F) removes the residual phonon-sideband emission, while leaving the zero-phonon line nearly unaffected.
The photons are then coupled to a single-mode fiber where, under $\pi$-pulse excitation, we measure $\SI{17.1}{MHz}$ of single-photons with an $85\%$ efficient detection system.
Under the same pumping conditions, lifetime measurements reveal a single-photon decay rate of $\tau_l \sim \SI{204}{ps}$ \newline

\begin{figure*}[h]
    \centering
    \includegraphics[width=0.8\textwidth]{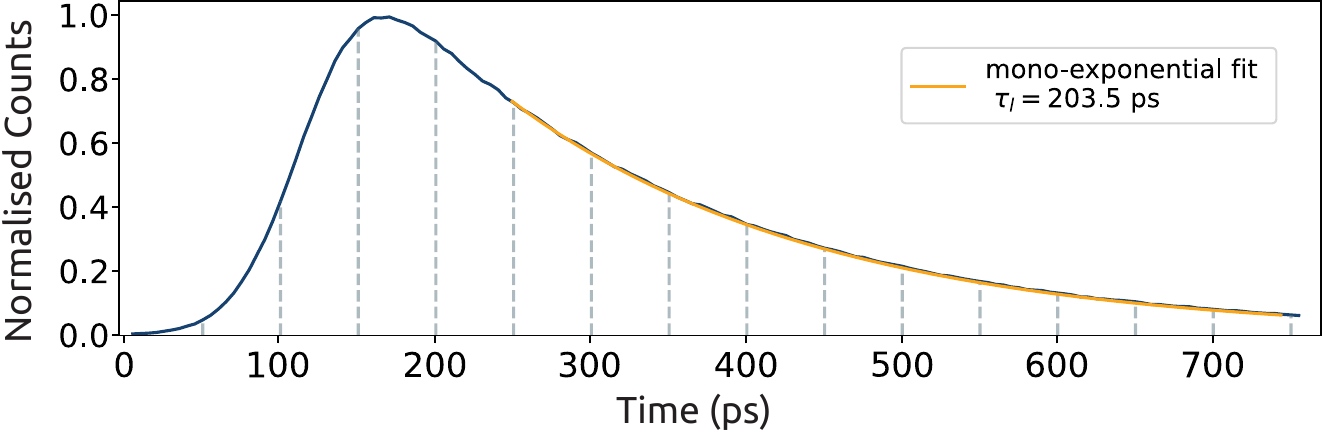}\vspace{0mm}	
    \caption{\textbf{Time resolved fluorescence.} Correlation histogram between single-photon signal and laser clock at $\pi$ pulse excitation. The mono-exponential fit function yields a decay time around $\SI{204}{ps}$.}
    \label{fig:lifetime}
 \end{figure*}


\section{Correlation histograms \label{sec_SM:auto-correlation}}
The temporal $m-$th order correlation functions integrate over a period $T$ is defined as
\begin{equation}
\label{eq_SM:gm}
    g^{(m)}(\tau_1, ..., \tau_{m-1}) =  \int_0^T dt
          \frac{\langle \hat{a}^\dagger(t) \hat{a}^\dagger(t+\tau_1) ... \hat{a}^\dagger(t+\tau_{m-1}) \hat{a}(t+\tau_{m-1}) ... \hat{a}(t+\tau_{m-1}) \hat{a}(t) \rangle}
          { \langle \hat{n}(t) \rangle \langle  \hat{n}(t + \tau_1) \rangle ... \langle  \hat{n}(t + \tau_{m-1}) \rangle } \ ,
\end{equation}
with $\hat{a} (\hat{a}^\dagger)$ the annihilation (creation) operator of a single mode of the field, $\hat{n}(t) = \hat{a}^\dagger(t) \hat{a}(t) $ the number operator, and  $\mathbf{\tau} = \{ \tau_1, ..., \tau_{m-1} \}$ a vector of time delays. 
The connection of this quantity with the photon number probabilities made it an essential tool in quantum optics. In particular, using the commutators for bosonic field operators $[\hat{a}(t), \hat{a}^\dagger(t')] = \delta_{t, t'}$ we can rewrite eq. \ref{eq_SM:gm} as follows
\begin{equation}
g^{(m)}(0, ..., 0) = \frac{\langle \hat{n} (\hat{n}-1) ... (\hat{n} - m + 1)\rangle}{ \langle \hat{n} \rangle^m} \ ,
\end{equation}
where the time dependency from $t$ has been dropped for simplicity. The expectations in this formula are then expressed in terms of the photon number probabilities  $\{p_n\}_{\mathbb{N}}$ as 
\begin{equation}
    \langle {\hat{n} (\hat{n}-1) ... (\hat{n} - m + 1)} \rangle  = \sum_{n = m}^{\infty} n (n-1) ... (n-m+1) p_n \ , \hspace{1 cm}
    \langle \hat{n} \rangle = \sum_{n = 1}^{\infty} n p_n \ .
\end{equation}
The details on how to retrieve the photon number probabilities are given in Sec. \ref{sec_SM:pn-probabilities}, therefore we will now focus on the measurement of the correlation functions and their graphical understanding. 
\subsection{Hanbury Brown-Twiss experiment}

In order to measure the $g^{(m)} (\mathbf{\tau}) $ it is possible to generalise the Hanbury Brown-Twiss experiment \cite{Brown1956}, as sketched in Fig. 1 for the case $m=4$. 
An input field is equally split in $m$ different paths with the help of beam splitters and the outputs are measured with detectors and a time-tagger that records the arrival time of each click and correlates the events as shown in Fig. \ref{fig_SM:correlations-scheme}. 

While our treatment focuses on a source under pulsed excitation, the same principles apply to continuous-wave operation. In both cases, the timeline is divided into discrete bins with a specific bin width. For continuous-wave excitation, all bins contribute to the analysis of source properties. In contrast, under pulsed excitation, only bins corresponding to signal peaks are relevant, and background counts between peaks can, depending on the application, often be neglected.
\newline
The correlation histogram reflects the $(m-1)$ time delays between the $m$ channels in the measurement. The simplest patterns arise in uncorrelated bins, where $\tau_1 \ne \tau_2 \ne ... \ne \tau_{m-1} \ne 0 $. Counts in these bins are proportional to the product of the count rates at individual detectors, representing the performance of the photonic source without collective phenomena. These uncorrelated counts are used to normalize the histogram, rendering the correlation data independent of linear losses.\newline 
The remaining bins capture multi-photon emission phenomena. Below, we discuss specific cases before generalizing to higher-order correlations.

\subsubsection{The case $m=2$}
For $m=2$ two channels are correlated, producing a one-dimensional histogram as shown in Fig.~\ref{fig_SM:correlations-scheme}. Uncorrelated peaks occur at  $\tau \ne 0$ and are proportional to the product of the single-detector counts, $H^{(2)}(\tau) = C_1 C_2$. The central peak at  $\tau = 0$ corresponds to coincidence counts, $H^{(2)}(0) = C_{12}$.
 The second-order correlation function at zero time delay is calculated as: \begin{equation} g^{(2)}(0) = \frac{H^{(2)}(0)}{\langle H^{(2)}(\tau) \rangle} \hspace{0.5 cm} \text{for} \hspace{0.5 cm} \tau \neq 0 \ , \label{eq_SM:g2-normalization} 
 \end{equation} 
by dividing the coincidence counts by the average uncorrelated counts.

\subsubsection{The case $m=3$}
\begin{figure*}[t]
    \centering
    \includegraphics[width=0.8\textwidth]{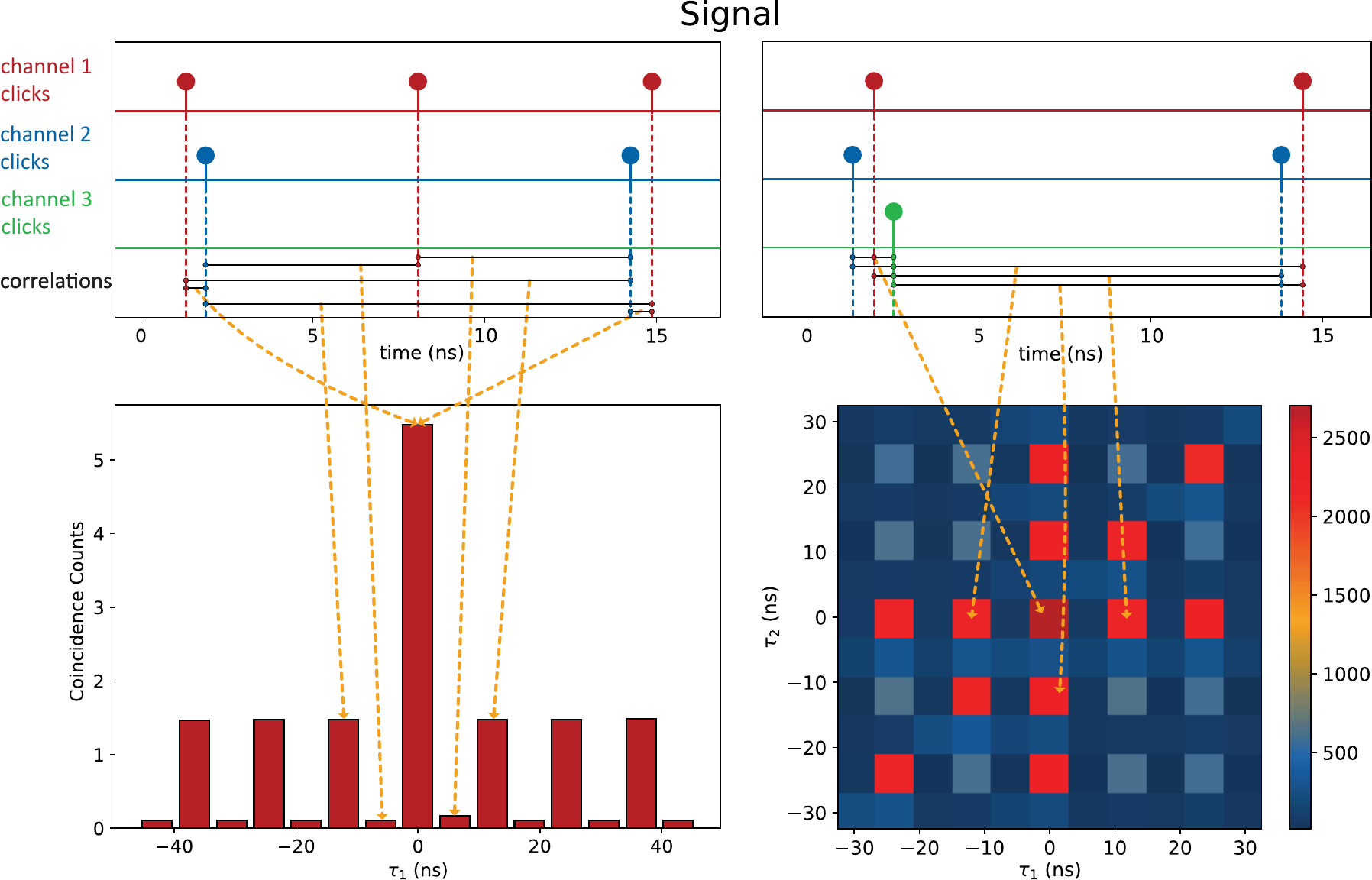}\vspace{0mm}	
    \caption{\textbf{Building correlation histograms.} Scheme for constructing multi-start multi-stop correlation histograms from the signal recorded by the time tagger. To build the histogram shown on the left, the time difference between each signal recorded in channel 1 and each signal recorded in channel 2 is calculated. As a result, every pair of clicks generates a count in the histogram at $\tau_1 = t_2 - t_1$ where $t_i$ is the arrival time of a photon in the $i$-th channel. The situation depicted on the right follows an analogous procedure. In this case, each combination of clicks across the three channels produces a count in the two-dimensional histogram, determined by the coordinates $\tau_2 = t_3 - t_2$ and $\tau_1$.}
    \label{fig_SM:correlations-scheme}
 \end{figure*}

For $m=3$, three channels are correlated, producing a two-dimensional histogram with more intricate patterns. 
Uncorrelated peaks, where $\tau_1 \ne \tau_2 \ne 0$, are proportional to the product of single-detector counts, $H^{(3)}(\tau_1, \tau_2) = C_1 C_2 C_3$.
The central peak at $\tau_1 = \tau_2 = 0$ corresponds to three-fold-coincidence counts, $H^{(3)}(0, 0) = C_{123}$.
Additionally, there are regions corresponding to two-fold coincidences, such as $H^{(3)}(\tau, \tau) = C_1 C_{23}$ for $\tau \ne 0$.
When normalized to the uncorrelated counts as in Eq. \ref{eq_SM:g2-normalization} we estimate the $g^{(3)} (0, 0)$ from the central peak and the second order correlation along the following lines 
\begin{equation}
    g^{(3)} (\tau, \tau) = g^{(3)} (\tau, 0) = g^{(3)} (0, \tau) = g^{(2)} (0) \ , \hspace{0.5 cm} \text{for} \hspace{0.5 cm}  \tau \neq 0 \ ,
\end{equation}
as shown in Fig. \ref{fig_SM:correlations-scheme}.

\subsubsection{The case $m=4$ and higher}
For $m=4$, the four-dimensional histogram contains estimates for the $g^{(4)}(0,0,0)$ as well as lower-order correlations. For example, $ H^{(4)}(\tau, 0, 0)  = C_2 C_{134}$ for $\tau \ne 0$, implying that $ g^{(4)}(\tau, 0, 0) = g^{(3)} (0, 0)$.
Additionally, patterns emerge corresponding to pairwise coincidences between four channels, such as $H^{(4)}(0, \tau, \tau) = C_{12} C_{34}$ for $\tau \ne 0$. These patterns appear as lines in Fig.~2e, and mathematically satisfy: 
\begin{equation} g^{(4)}(0, \tau, \tau) = g^{(4)}(\tau, 0, \tau) = g^{(4)}(\tau, \tau, 0) = \left[ g^{(2)}(0) \right]^2 \ , \hspace{0.5 cm} \text{for} \hspace{0.5 cm} \tau \neq 0 \ . 
\end{equation}
The analysis extends naturally to higher-order correlations, where each peak of the histogram corresponds to coincidence clicks involving $1\le k\le m$ channels, allowing the estimation of $g^{(k)}(0)$.

\subsection{Normalisation and error calculation}
Practically, the $m$-th order correlation function at zero time delay is estimated with $1\sigma$ confidence levels ($CL$) as 
\begin{equation}
    \label{eq_SM:gn}
    g^{(m)}(0) = \frac{N_\text{c}}{N_\text{u}} \pm \sqrt{\left(\frac{\delta N_\text{c} }{N_\text{u}}\right)^2 + \left(\frac{N_\text{c} \delta N_\text{u}}{N_\text{u}^2} \right)^2 }  \ ,  
\end{equation}
where $N_\text{c}$ are the integrated counts at zero time-delay, $N_\text{u}$ is the average count in uncorrelated peaks, and $\delta N_\text{c}, \delta N_\text{u}$ are their respective uncertainties. For every peak, the integration window is set to $\SI{3}{ns}$.

The uncertainty $\delta N_\text{u}$ is computed as the standard deviation of uncorrelated peaks, while $\delta N_\text{c}$ is estimated using $CL = 68.3\%$  for a Poisson-distributed variable.
For $N_c > 100$, the Gaussian approximation $\delta N_\text{c} = \sqrt{N_\text{c}}$ is used. For lower counts, the exact Poisson confidence intervals are calculated using (see Eq. (33.59) in Ref. \cite{Nakamura2010}): 
\begin{equation}
    \delta N_\text{c}^\text{up} = \frac{1}{2} Q_{\chi^2}\!\left(\frac{\alpha}{2}; 2N_\text{c}\right) - N_\text{c} \ , \hspace{1 cm}
    \delta N_\text{c}^\text{low} = N_\text{c} - \frac{1}{2} Q_{\chi^2}\!\left(1 - \frac{\alpha}{2}; 2(N_\text{c} + 1)\right)
\end{equation}
where $Q$ is the chi-square quantile, and $\alpha = 1 - CL$. The uncertainties are propagated independently to compute the result in \ref{eq_SM:gn}.

\section{Photon number probabilities \label{sec_SM:pn-probabilities}}
\subsection{Extraction of atomic probabilities from measured data}
As described in Section IIB of the main text, we calculate the brightness at the detectors, $B' = \sum_{n=0}^4 p_n'$, by summing the count rates of all four output channels, avoiding multiple counts in case of coincidence clicks, and dividing by the repetition rate.
Then, we extract the photon number probabilities solving the following system of equations
\begin{equation}
    \begin{cases}
    g^{(2)}(0) = \frac{\langle n (n-1) \rangle}{\langle n\rangle^2} \ , \\
    g^{(3)}(0, 0) = \frac{\langle n (n-1) (n-2) \rangle}{\langle n  \rangle^3} \ , \\
    g^{(4)}(0, 0, 0) = \frac{\langle n (n-1) (n-2) (n-3)\rangle}{\langle n  \rangle^4} \ , \\
    B' = \sum_{n=0}^4 p_n'
    \end{cases}
\end{equation}
where $\langle n  \rangle = \sum_{n=0}^4 n p_n'$ and $\langle n (n-1) ... (n - i)  \rangle = \sum_{n=i}^4 n (n-1) ... (n - i) p_n'$.
Further, we introduce a loss parameter $\eta_{\text{t}}$, expressing the probability that an emitted photon causes a click of the detector, implicitly assuming that every photon has the same probability of being lost - whether it comes as part of a multi-photon bundle or as a pure single-photon. In particular, the relation between the emitted and detected signal is mathematically expressed by the following equations
\begin{align}
    \label{eq_SM:loss}
    p_0' &= p_0 + (1 - \eta_{\text{t}}) p_1 + (1 - \eta_{\text{t}})^2 p_2 + (1 - \eta_{\text{t}})^3p_3 + (1 - \eta_{\text{t}})^4p_4 \nonumber \ ,\\
    p_1'&= \eta_{\text{t}} p_1 + 2\eta_{\text{t}}(1 - \eta_{\text{t}})p_2 + 3\eta_{\text{t}}(1 - \eta_{\text{t}})^2p_3 + 4\eta_{\text{t}}(1 - \eta_{\text{t}})^3p_4 \nonumber \ ,\\
    p_2' &= \eta_{\text{t}}^2p_2 + 3\eta_{\text{t}}^2(1 - \eta_{\text{t}})p_3 + 6\eta_{\text{t}}^2(1 - \eta_{\text{t}})^2p_4 \ , \\
    p_3' &= \eta_{\text{t}}^3 p_3 + 4\eta_{\text{t}}^3(1 - \eta_{\text{t}})p_4 \nonumber \ ,\\
    p_4' &= \eta_{\text{t}}^4 p_4 \ .  \nonumber
\end{align}
Finally, we assume that the QD produces at least photon per incident $\pi-$pulse, hence $B_{\pi} = \sum_{n=0}^4 p_n = 1$.
Exploiting this relation we invert Eqs. \ref{eq_SM:loss} and extract the overall transmission probability as $\eta_{\text{t}} \simeq 0.25$.
This allows us to retrieve the photon number probabilities as emitted by the QD $\{p_n\}$ for each excitation power, where the vacuum probability is derived by the normalization rule as $p_0 = 1 - B$. \newline

\subsection{Bunching auto-correlation and multi-photon component}
\begin{figure*}[ht]
    \centering
    \includegraphics[width=0.4\textwidth]{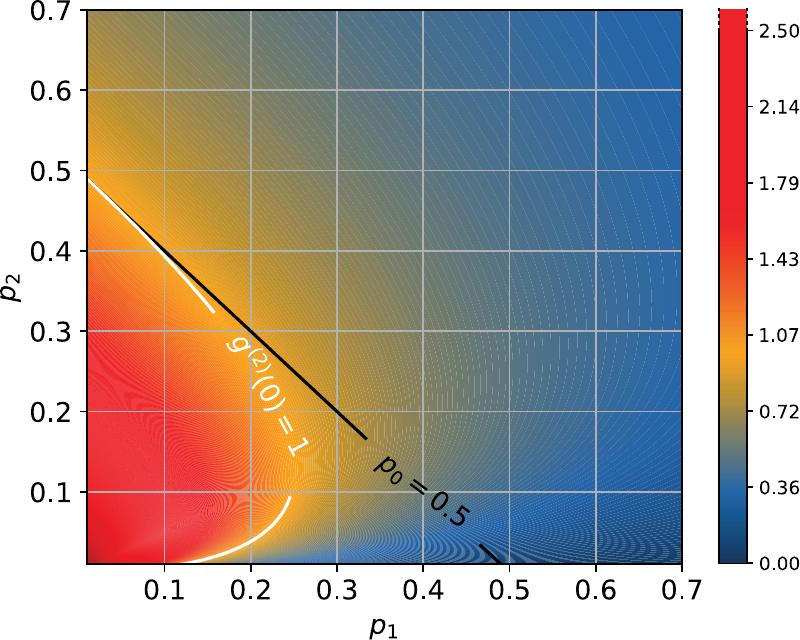}\vspace{0mm}
    \caption{\label{fig:g2} \textbf{Dependency of the autocorrelation function on photon number probabilities.} The graph highlights the importance of the vacuum probability in order to obtain a bunching auto-correlation.}
    
\end{figure*}
Let us consider a light source that displays a bunching second-order correlation histogram, as shown in Fig. 2b. To obtain bunching statistics, the auto-correlation function must satisfy $g^{(2)}(0) > 1$.
Given that auto-correlation functions are largely loss independent, we can equivalently express them as a function of the photon number probabilities at source level $p_n$

\begin{equation}
    g^{(2)}(0) = \frac{2p_2}{(p_1 + 2p_2)^2} .
\end{equation}

For simplicity, we neglected states containing more than two photons, that is $p_{n>2} = 0$.
Interestingly, we then find that the condition $g^{(2)}(0) > 1$ does \textit{not} imply $p_2 > p_1$, as shown in Fig. \ref{fig:g2}.
We emphasize that the vacuum probability plays a crucial role; in this simplified picture, if $p_0 \le 0.5$ the correlation histogram can not display bunching statistics, regardless of the other probability values. 

This concept can be readily extended to higher orders autocorrelation functions, where $g^{(m)}(0) > 1$ does not imply the prevalence of the $p_m$ component in the output state.

\bibliography{supplementary}